\documentclass{PoS}

\newcommand{\bmat}{\left ( \begin{array}{cc}}
\newcommand{\emat}{\end{array} \right )}
\newcommand{\vect}{\left ( \begin{array}{c}}
\newcommand{\evect}{\end{array} \right )}
\newcommand{\bitem}{\begin{itemize}}
\newcommand{\eitem}{\end{itemize}}

\newcommand{\be}{ \begin{eqnarray}}
\newcommand{\ee}{\end{eqnarray} }

\newcommand\nn{\nonumber}
\newcommand{\Tr}{{\rm Tr}}
\newcommand{\dl}{   $ }
\newcommand{\dr}{ $   }

\title{Discretization Effects in the $\epsilon$ Domain of QCD}

\ShortTitle{Discretization Effects }

\author{Mario Kieburg\\
        Fakult\"at f\"ur Physik, Postfach 100131, 33501 Bielefeld, Germany\\
        E-mail: \email{mkieburg@physik.uni-bielefeld.de}}
\author{K. Splittorff\\
       Discovery Center, The Niels Bohr Institute \\
       University of Copenhagen, \\
       Blegdamsvej 17, DK-2100 Copenhagen, Denmark\\
       E-mail: \email{split@nbi.dk}}

\author{\speaker{Jacobus J. M. Verbaarschot}\\
        Department of Physics and Astronomy, State University of New York at Stony Brook, NY 11794-3800, USA\\
        E-mail: \email{jacobus.verbaarschot@stonybrook.edu}}
 
\author{Savvas Zafeiropoulos\\
        Department of Physics and Astronomy, State University of New York at Stony Brook, NY 11794-3800, USA and\\
        Laboratoire de Physique Corpusculaire, Universit\'e Blaise Pascal, CNRS/IN2P3 
63177 Aubi\`ere Cedex, France\\
        E-mail: \email{zafeiropoulos@clermont.in2p3.fr}}

\abstract{At nonzero lattice spacing the QCD partition function with Wilson quarks undergoes either a second order phase transition to the Aoki phase for decreasing quark mass or shows a first order jump when the quark mass changes sign. 
We discuss these phase transitions in terms of Wilson Dirac spectra and show that the first order scenario can only occur in the presence of dynamical quarks while in the quenched case we can only have a transition to the Aoki phase. The exact microscopic spectral density of the non-Hermitian Wilson Dirac operator with dynamical quarks is discussed as well. We conclude with some remarks on discretization effects for the overlap Dirac operator.}

\FullConference{31st International Symposium on Lattice Field Theory - LATTICE 2013\\
		July 29 - August 3, 2013\\
		Mainz, Germany}

\begin{document}

\section{Introduction}

The Wilson term violates the anti-Hermiticity of the Dirac operator
so that its eigenvalues will be scattered into the complex plane. When the
quark mass is inside the strip of eigenvalues the partition function is in
a phase known as the Aoki phase \cite{aoki}, and a reliable
extrapolation to the continuum limit is not possible. This scenario
happens in the quenched case for sufficiently small quark masses
and temperature 
\cite{KSV}.
For dynamical quarks, a different scenario is possible. Because of the
fermion determinant, the eigenvalues will be repelled from the quark mass and
when the quark mass crosses zero a collective jump of the 
eigenvalues  to the other side of the imaginary axis can take 
place \cite{KSV}. This is known as the first order or Sharpe-Singleton scenario 
\cite{sharpe,Golterman}, which was first
discovered by means of Wilson chiral perturbation theory.
In this scenario there is no transition to a different phase
and extrapolation to the continuum limit is possible for all values
of the quark mass, $m$, and of the lattice spacing, $a$.
What happens can be analyzed by means of Wilson chiral perturbation
theory. The terms in the chiral Lagrangian are determined by chiral symmetry in the same way as the terms involving the quark mass.
The lowest order term in $a$ can be absorbed by a redefinition
of the mass term leaving us with three terms of second order in $a$ with 
low-energy constants $W_{6/7/8}$.  
The  spectrum of the Wilson Dirac operator can be obtained
by extending the chiral Lagrangian with fermionic and bosonic 
valence quarks \cite{Sharpe,NS,Golterman,DSV}. In the microscopic limit it can also be  obtained  by analyzing the equivalent random matrix
theory (RMT) \cite{DSV, KVZ,nagao,KVZ-large}. Among others this 
  provides us with simple relations
to determine the low-energy constants \cite{KVZ-large}. One important 
effect of a non-zero lattice spacing is the broadening of the zero modes to
a finite width $\sim a/\sqrt V$ (for small $a$) \cite{DSV} as was first observed in \cite{Luscher}. Other analytical predictions for the Wilson
Dirac spectra have also been confirmed  by lattice simulations
\cite{deuzeman,heller1,heller2}.

Because of the absence of an exact chiral symmetry it is hard to approach the chiral
limit with Wilson fermions, but this is possible by means of the overlap 
Dirac operator \cite{neuberger}
or by domain wall fermions \cite{kaplan}.
Since the overlap Dirac operator  
is based on the Wilson Dirac operator,
we expect that it will inherit a remnant of the discretization effects.
In particular, if the Wilson Dirac operator is in the Aoki phase, one may 
question if the overlap operator will not describe the continuum limit
of lattice QCD. We investigate this in the framework of a random
matrix model for the Wilson Dirac operator discussed in Sec.~\ref{sec2}. Moreover we discuss the relation between the spectra of the Wilson Dirac operator and the overlap Dirac operator in Sec.~\ref{sec3}.   

\section{Wilson Chiral Perturbation Theory}\label{sec2}

The Wilson Dirac operator has the following block structure
\be
D_{\rm W }= \bmat a A & iW \\i W^\dagger & aB \emat
\label{wilmat}
\ee
with $W$ a complex matrix and $A$ and $B$ Hermitian matrices. This operator is
$\gamma_5$-Hermitian ($D_{\rm W}^\dagger = \gamma_5 D_{\rm W} \gamma_5$) so that its
eigenvalues occur in complex conjugate pairs or are real. In the $\epsilon$ domain
of QCD ($V\to\infty$ while $mV$ and $a^2V$ are kept constant)
the chiral Lagrangian is given by
\cite{sharpe,RS,BRS,DSV}.
\be
-{\cal L} &=&\frac 12mV\Sigma \Tr (U+ U^\dagger) -\frac 12 zV\Sigma  \Tr (U- U^\dagger)\nn \\
&& \hspace*{-1cm}-a^2 V W_6 [{\rm Tr} (U+U^\dagger)]^2
-a^2 V W_7 [{\rm Tr} (U-U^\dagger)]^2
 -a^2 V W_8 \Tr( U^2 +U^{-2}).
\label{lmicro}
\ee
The trace squared $O(a^2)$ terms can be linearized by introducing a random mass
\cite{ADSV}. For $ W_6 <0$ we have
\be
\exp\left[-a^2 V W_6 \Tr^2( U +U^{-1})\right] =\int_{-\infty}^\infty \frac{dy}{\sqrt{16\pi V|W_6| a^2}} \exp\left[-\frac{y^2}{16 V|W_6| a^2} - \frac{y}{2}\Tr ( U +U^{-1})\right],
\ee
and the term proportional to  $W_7$  can be linearized as (for $W_7 <0 $) 
\be
\exp\left[-a^2 V W_7 \Tr^2( U -U^{-1})\right] =\int_{-\infty}^\infty \frac{dy}{\sqrt{16\pi V|W_7| a^2}} \exp\left[-\frac{y^2}{16 V|W_7| a^2} - \frac{y}{2}\Tr ( U -U^{-1})\right].
\ee
%while for $W_8>0 $ the following identity holds
%\be
%e^{-a^2 V W_8 \Tr( U^2 +U^{-2})} \sim \int d\sigma 
%e^{-\Tr(\sigma^2/(16 VW_8 a^2)) - \frac i2 \Tr (\sigma( U +U^{-1}))}.
%\nn 
%\ee
Therefore, the trace squared terms can be generated by a random mass for
the real part of the eigenvalues of $D_{\rm W}$ (for $W_6$) and a chiral random mass
for the eigenvalues of $D_5=\gamma_5 D_{\rm W}$ (for $W_7$). For the opposite sign
of $W_6$ and $W_7$, these terms cannot originate from eigenvalue fluctuations
\cite{KSV}.
%The random mass corresponding to $W_8$ cannot directly be
%interpreted in terms of collective eigenvalue fluctuations and in fact violates
%$\gamma_5$-Hermiticity.  
The sign of $W_8$ follows from positivity requirements
of the partition function \cite{ADSV} at fixed index $\nu$ of the Dirac operator yielding
\be
W_8 -W_6 -W_7 >0.
\ee
 Another inequality follows from the width of the strip 
parallel to the imaginary axis
forming the support of the eigenvalues of $D_{\rm W}$  which for small $a$ is proportional to $W_8 -2 W_6$ so that
\be
W_8-2W_6 > 0.
\ee 
Using partially quenched chiral perturbation theory, it can  additionally 
be shown
that $W_8>0$ independently of the value of $W_6$ and $W_7$ \cite{sharpe-hansen}.
In Table 1 we give some recent results \cite{Herdoiza} for the low energy constants
which are consistent with the inequalities given above.

\begin{table}[h!]
\centerline{\begin{tabular}{c|cc}
& \dl W_6'\dr & \dl W_8' \dr\\
\hline
Iwasaki &0.0049(38)&--0.0119(17)\\
tlSym& 0.0082(34) & --0.0138(22)
\end{tabular}}
\caption{Recent lattice results \cite{Herdoiza} for the low energy constants $W_6'=-W_6$ 
and $W_8'=-W_8$.}
\end{table}

It is instructive to interpret the first order scenario  in terms of
eigenvalues of the Dirac operator. Because of the random mass, due
to $W_6$, the strip of eigenvalues exhibits collective fluctuations. For dynamical
quarks, this strip is repelled from the quark mass. However, when the quark mass
crosses zero, the strip moves to the other side of the imaginary axis. A first
order transition takes place if its half-width is less than the 
position of its center. The half-width of the strip is equal to 
$8W_8a^2/\Sigma$. The distribution of the Gaussian random mass is modified by
the two flavor partition function (which 
is equal to $\exp (2\Sigma V|m-y|)$ in the mean field limit)
and is thus given by \cite{KSV} 
\be
\exp\left[-\frac{y^2\Sigma^2 V}{16 |W_6|a^2}\right ] \to \exp\left[-\frac{y^2\Sigma^2 V}{16| W_6|a^2} +2\Sigma V |m-y|\right],
\ee
which is a Gaussian centered at  $\pm 16 |W_6|a^2/\Sigma$. We thus
find a first order scenario if $8W_8a^2/\Sigma<16 |W_6|a^2/\Sigma$ and,
because of $W_6<0$, we have the condition $W_8+2W_6 <0$ for the first
order scenario to occur \cite{Golterman}. The role of the fermion determinant is essential
in this argument. Without it, the distribution  of the random mass will be a
Gaussian centered at zero so that a first order scenario cannot occur.
Indeed, in the quenched case only a transition to the Aoki phase has been
observed in lattice simulations see \cite{KSV,Splittorff:2012hz}.

\begin{figure}[t!]
\includegraphics[width=4.9cm, height=4.6cm]{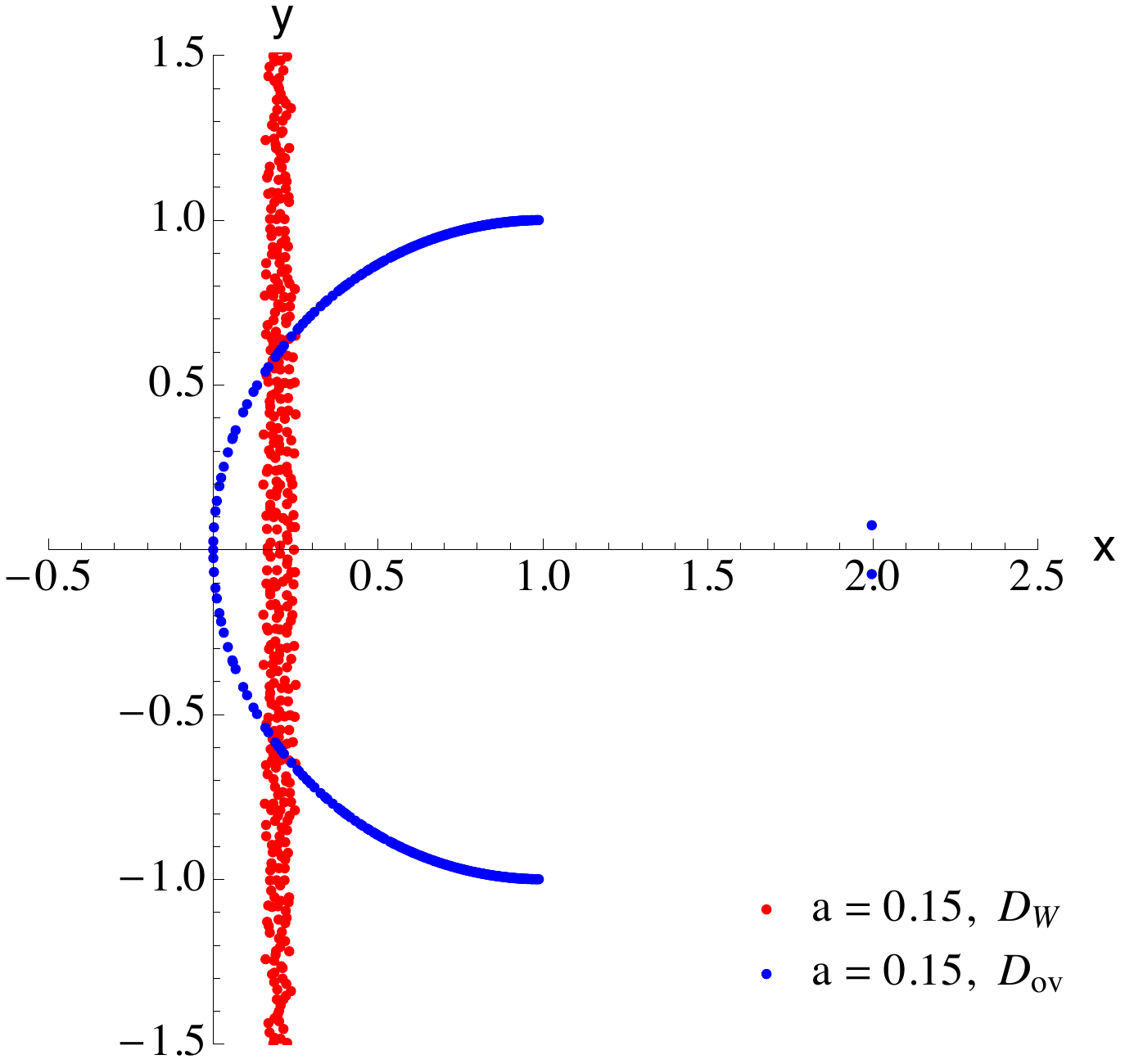}
\includegraphics[width=4.9cm, height=4.6cm]{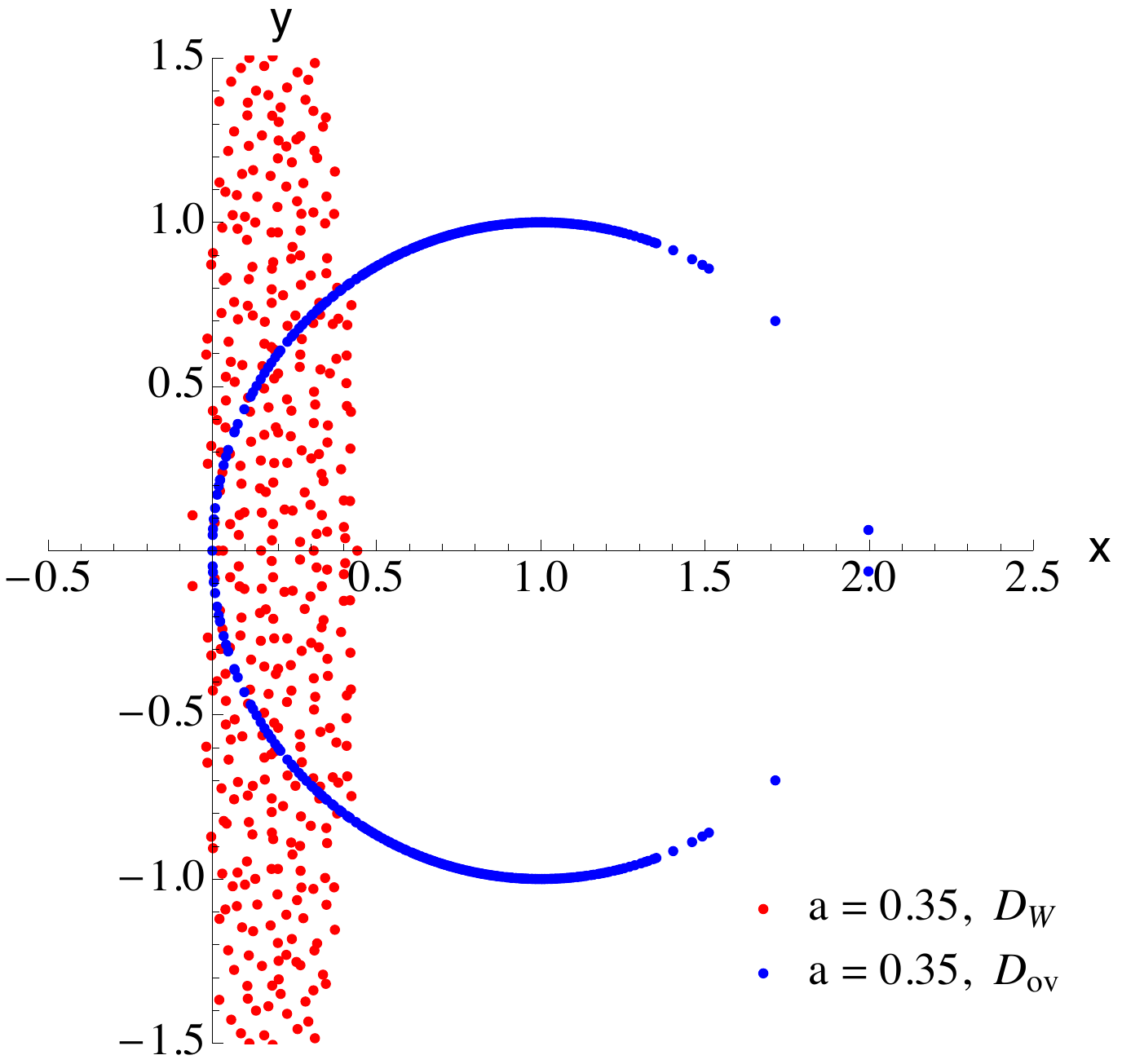}
\includegraphics[width=4.9cm, height=4.6cm]{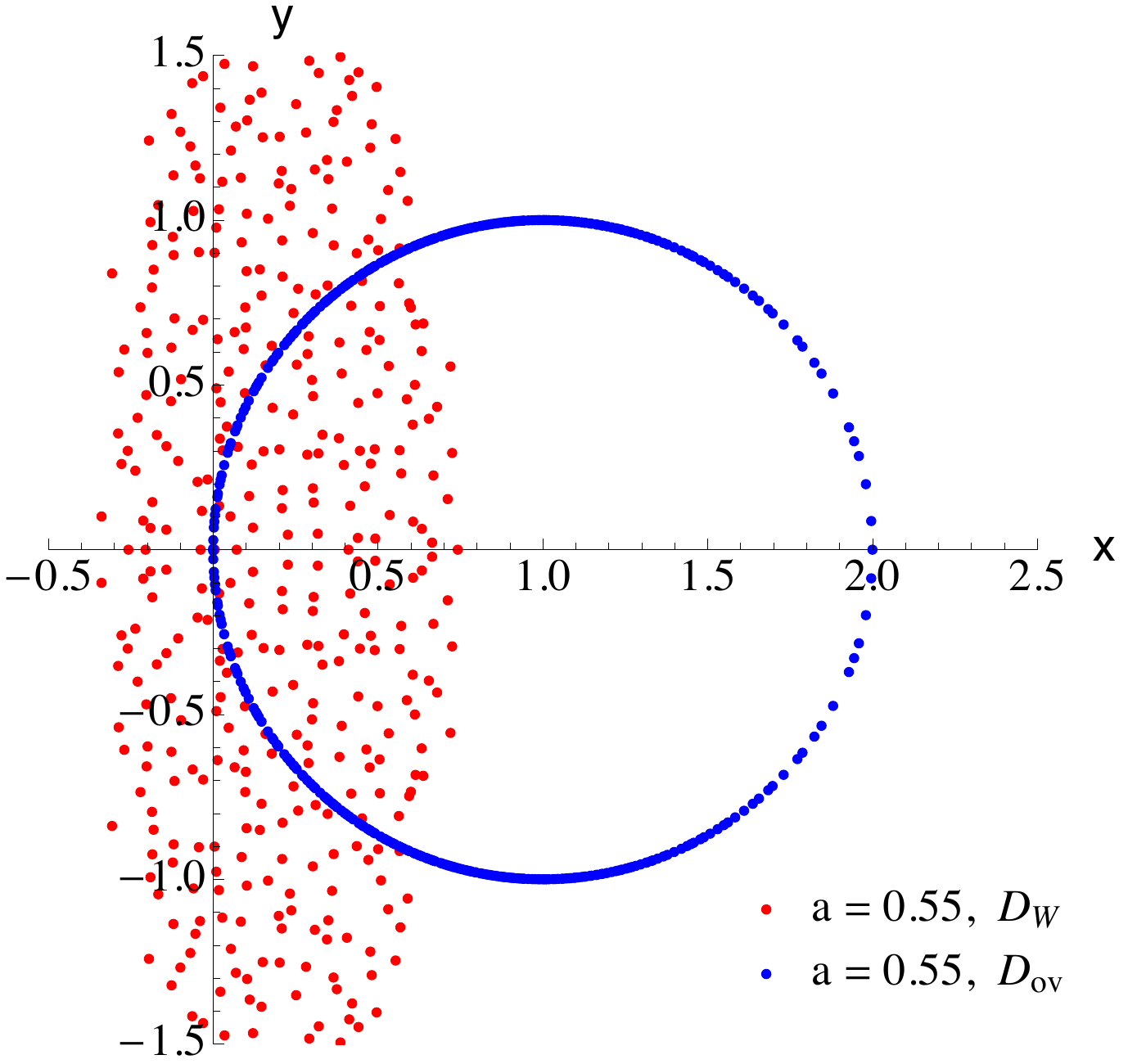}
\caption{Scatter plot of the eigenvalues of a single Wilson Dirac operator (red)
and its corresponding overlap Dirac operator (blue) with $m=0.2$ and different $a$ drawn from a random matrix ensemble with index $\nu=1$ and a $200\times 201$ off-diagonal block $W$.}
\label{fig1}
\end{figure}

\section{The Overlap Dirac Operator}\label{sec3}

In this section, we study the effect of the Wilson term on  spectra of the overlap Dirac
operator using the random matrix Wilson Dirac operator sharing the block structure 
of the Wilson Dirac operator~(\ref{wilmat}) whose matrix elements are replaced by independent Gaussian
random numbers. This RMT for the Wilson Dirac operator
yields the equivalent microscopic Lagrangian~(\ref{lmicro}).
The corresponding
massless overlap Dirac operator is defined in the usual way,
\cite{neuberger,Narayanan:1993sk,Narayanan:1993ss,Narayanan:1994gw, Edwards,split-andy}
\be \label{overlapdef}
D_{\rm ov} = 1 -\gamma_5 U {\rm sign}(\Lambda_5) U^{-1}, \qquad D_5 = (D_{\rm W}+ m)\gamma_5=U\Lambda_5U^{-1},
\ee
where $U$ is the unitary matrix that diagonalizes $D_5$ to its eigenvalues $\Lambda_5$.

\begin{figure}[t!]
\includegraphics[width=7.5cm, height=5cm]{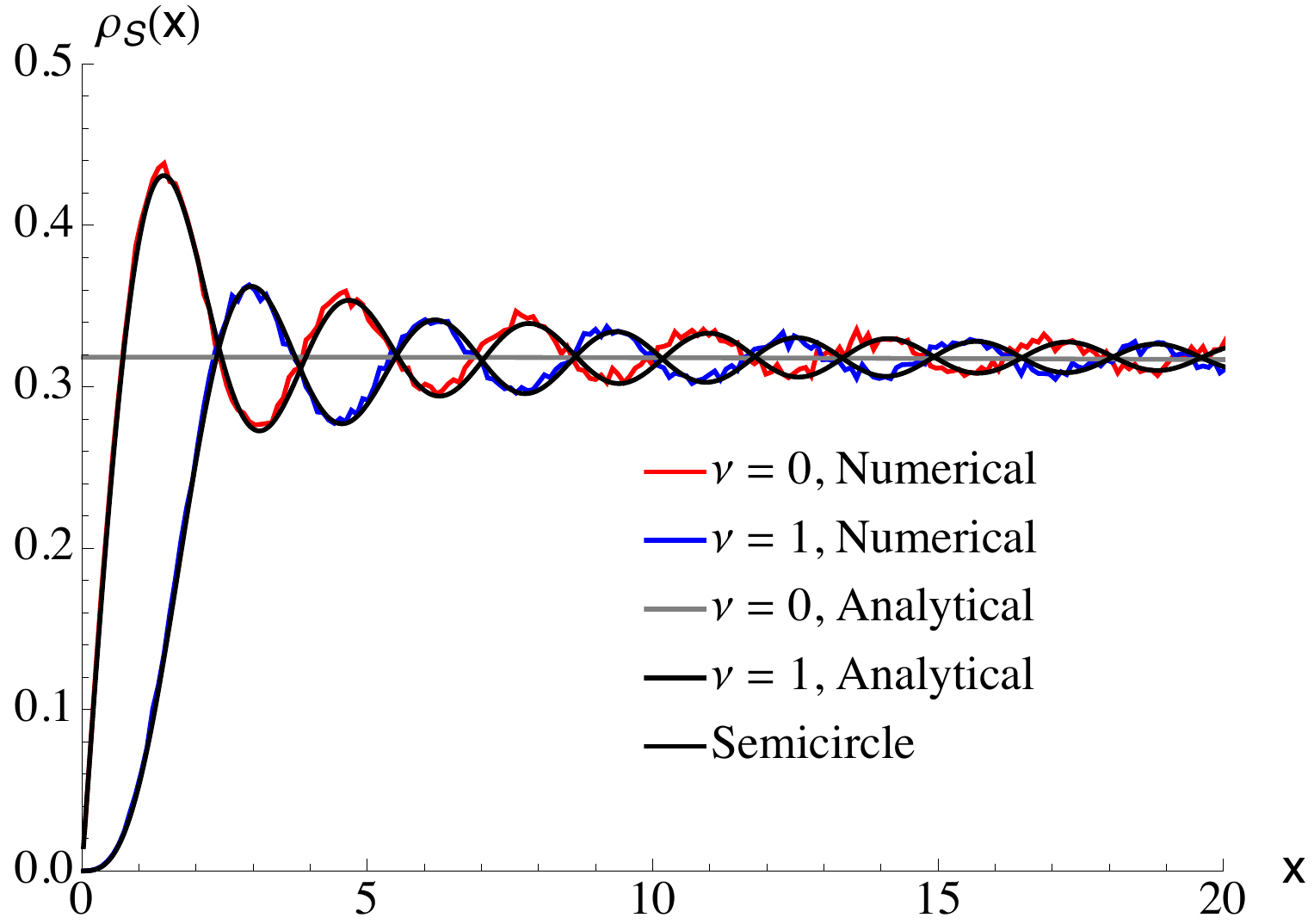}
\includegraphics[width=7.5cm, height=5cm]{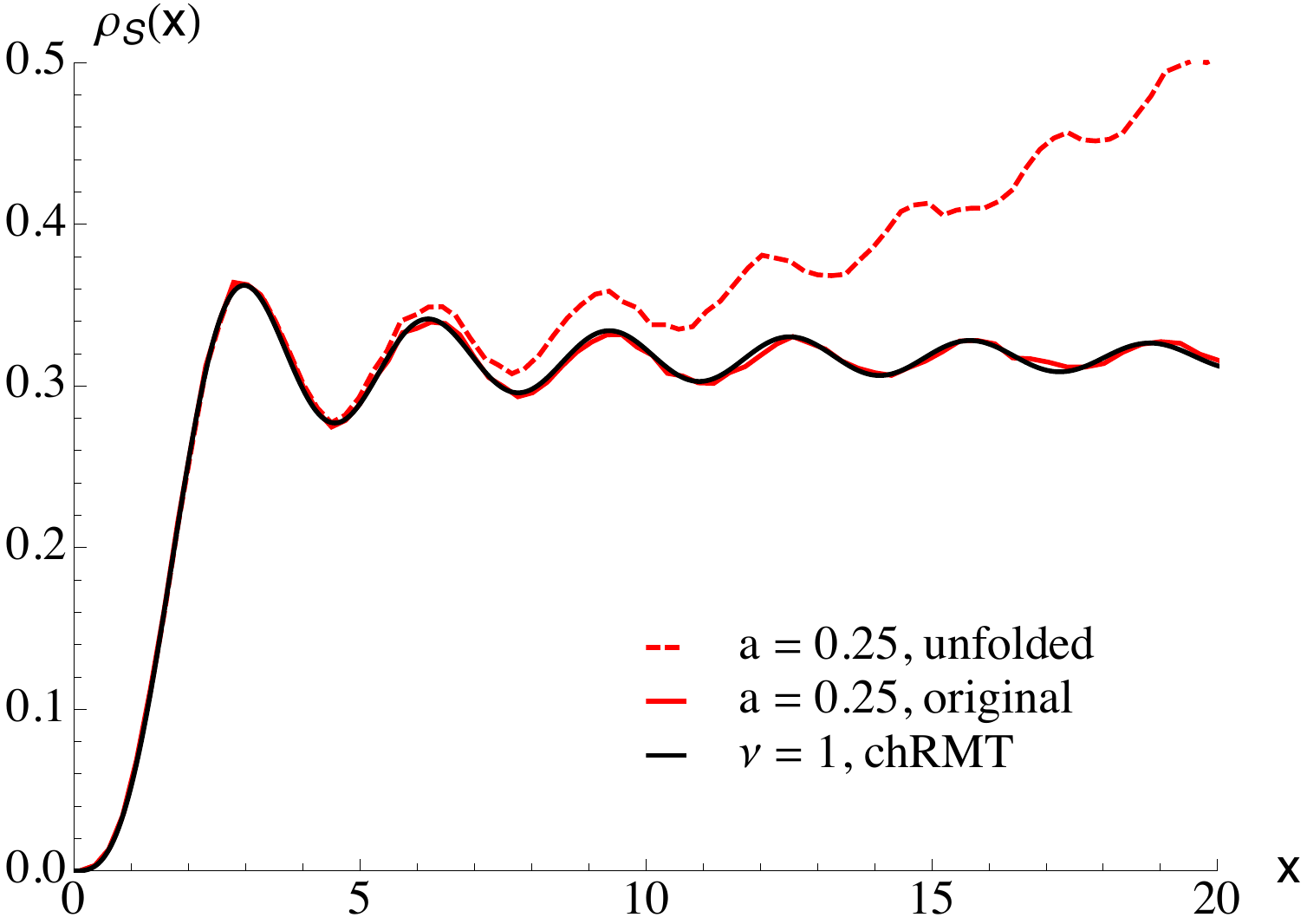}\hfill
%\centerline{\includegraphics[width=7cm,bb=200 500 400 750, clip=]{ov-a0.eps}}
\caption{\small 
{\bf Left:} The spectral density of the projected overlap Dirac operator obtained from a Wilson 
kernel with a large mass $m=100$, $a=0.3$ and index 
$\nu=0$ and $\nu=1$. The black curve
shows the analytical continuum result and the red and blue curve show 
the result from the
computed eigenvalues. The grey curve is the mean field result for the spectral 
density which is a semicircle with height $1/\pi$ at the origin.
{\bf Right:} Comparison of the spectral density of the overlap Dirac operator
for $a=0.15$ and $m=0.2$ (Red)  before (dashed) and after unfolding (full)
and   the analytical continuum chiral RMT result (black)  \cite{SV}.  
}
\label{fig2}
\end{figure}

It  looks drastic to replace the eigenvalues by their sign, but at vanishing
lattice spacing this does not change the theory. 
The reason is that eigenvectors contain the information on the eigenvalues. For $a=0$, the
Dirac operator $D_5$ can be block-diagonalized as
\be
D_5 = \bmat u & 0\\ 0& v \emat \bmat m &\lambda_k \\ \lambda_k & -m \emat
\bmat u^{-1} & 0 \\ 0 & v^{-1} \emat
\ee
with $u$ and $v$ unitary. Performing  an additional rotation by 
$ \tan(2\phi_k) = \lambda_k/m $ results in
\be
 \bmat m &\lambda_k \\ \lambda_k & -m \emat = 
\bmat \cos \phi_k & -\sin \phi_k \\ \sin \phi_k &\cos \phi_k \emat 
\bmat \sqrt{\lambda_k^2 +m ^2} & 0 \\ 0&- \sqrt{\lambda_k^2 +m ^2}  \emat
\bmat \cos \phi_k & \sin \phi_k \\ -\sin \phi_k &\cos \phi_k \emat .
\label{3.3}
\ee
After projection onto the sign of the eigenvalues, this rotation 
reproduces the left-hand of Eq. (\ref{3.3}) rescaled by a factor
$1/\sqrt{\lambda_k^2+m^2}$
so that the eigenvalues of the overlap Dirac operator are given by
\be
\lambda_{\rm ov, k}=1 - \frac {m\pm i\lambda_k}{\sqrt{\lambda_k^2+m^2}} = 1- e^{\pm 2i\phi_k}.
\ee
The stereographic projection of the eigenvalues onto the imaginary axis 
is given by
\be
\lambda_{\rm p, k} = \frac {\lambda_{\rm ov, k}}{1-\lambda_{\rm ov,k}/2} = \mp2i\tan \phi_k=\mp2 i\frac{\lambda_k}{m+\sqrt{\lambda_k^2+m^2}},
\ee
which for large mass, $m$, 
simplifies to
\be
\lambda_{\rm p, k} \approx\mp i\frac{\lambda_k}m + O\left (\frac {\lambda_k^2}{m^2} \right).
\ee
 At nonzero lattice spacing the overlap operator preserves a new 
form of chiral symmetry \cite{Luscher-gw} so that the
 expected pattern of chiral 
symmetry breaking  and eigenvalue  correlations are the same as in the continuum limit. 
In Fig.~\ref{fig1} we show scatter plots of eigenvalues of the random matrix  
Wilson Dirac operator and the corresponding overlap Dirac operator. 
For large Wilson mass, the eigenvalues move close to zero, and after 
rescaling we indeed find close agreement with the continuum chRMT result 
(see Fig.~\ref{fig2}).

For smaller Wilson mass or larger lattice spacing the eigenvalues of the overlap Dirac operator cover an increasing part of the unit circle, and in the Aoki phase,
the spectral density becomes  nonvanishing everywhere on the unit circle. 
In the normal phase the real eigenvalues of the overlap operator are necessarily
located at zero while in the Aoki phase they can be either at $\lambda =0 $ 
or $\lambda =2$ (see Fig. \ref{fig1}). As a consequence, the number of zero modes which determines
the microscopic spectral density
is 
not  equal to the index of the Wilson kernel and
the spectral density is a superposition of the spectra
with a different number of zero modes.  
This redistribution of the index of the Wilson kernel
may affect the $\theta$-dependence of the overlap partition function
which will be studied elsewhere.

\begin{figure}[t!]
\includegraphics[width=7.5cm, height=5cm]{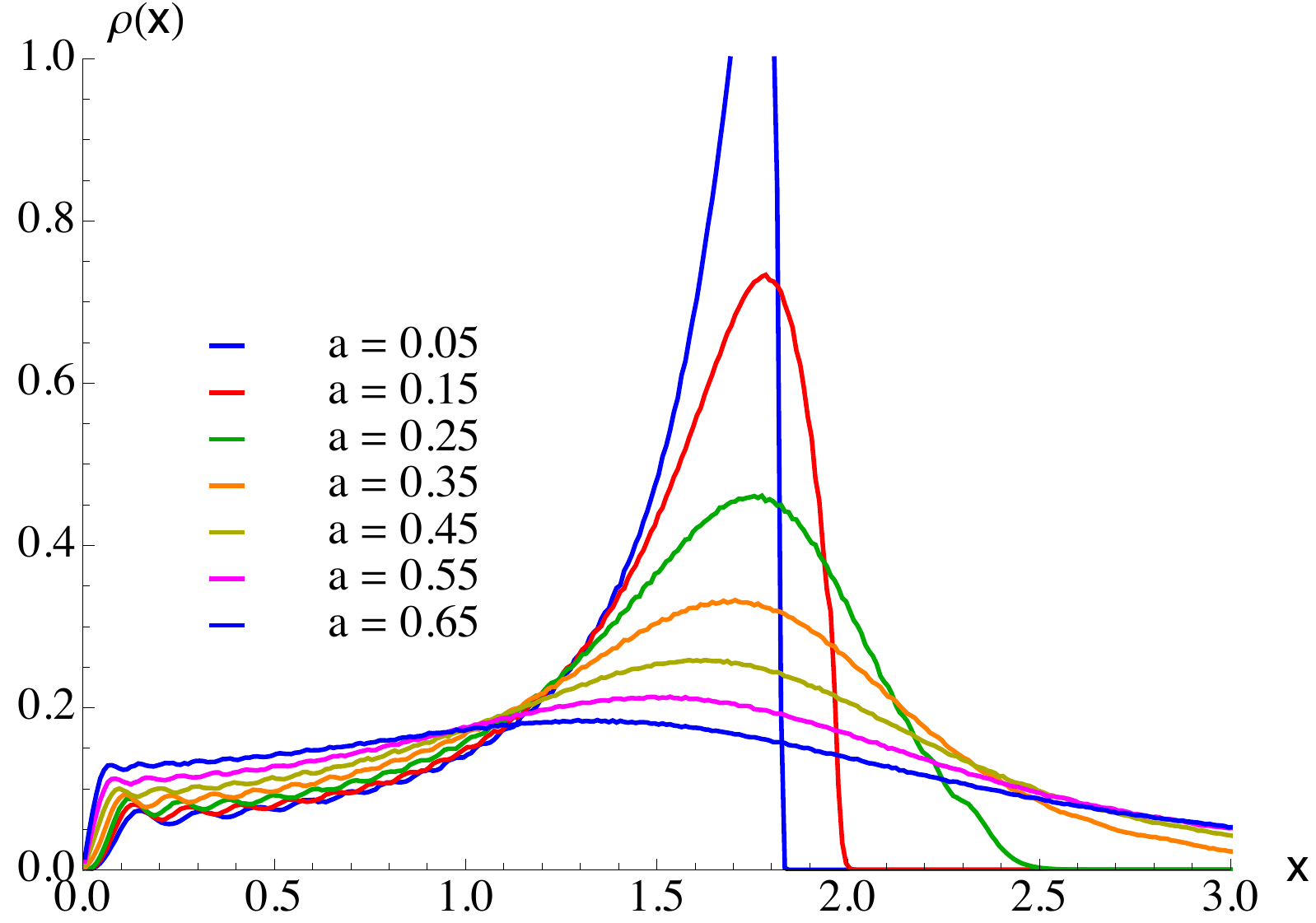}\hfill\includegraphics[width=7.5cm, height=5cm]{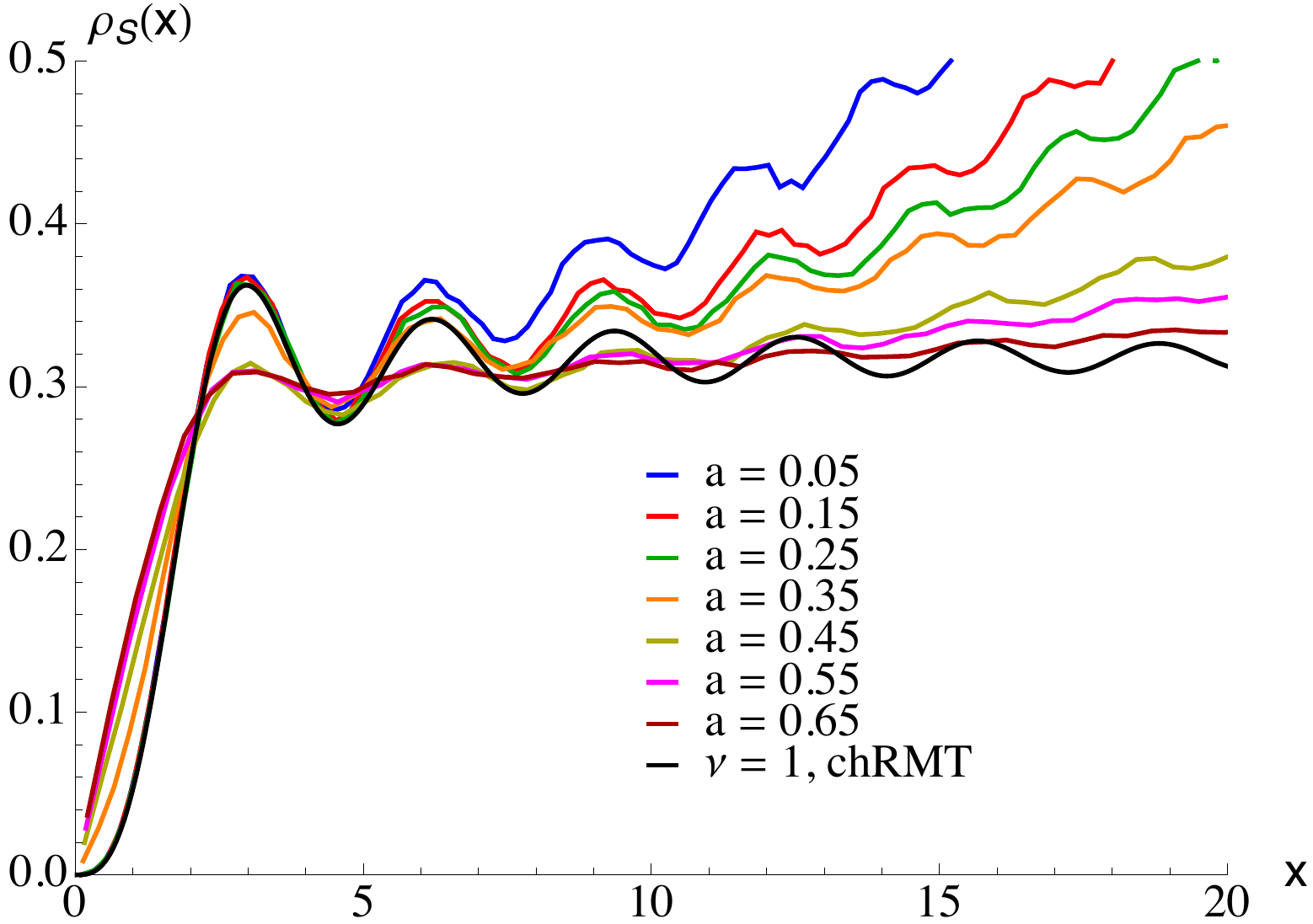}
\caption{\small The spectral density of the projected overlap Dirac 
spectra for  Wilson kernel with index $\nu=1$ and a small kernel mass $m=0.2$ 
and  $a=0.05$, $a=0.15$, $a=0.25$, $a=0.35$, $a=0.45$, $a=0.55$, $a=0.65$
(from top to bottom). The critical value for the transition to the
Aoki phase is $a=0.35$ (green curve). In the right figure the curves have
been rescaled, but otherwise the data are the same.}
\label{fig3}
\end{figure}

 The spectral behavior of $D_{\rm ov}$
is illustrated 
 by simulations of  the overlap operator with 
the Wilson RMT kernel, see Eq.~(\ref{wilmat}).
 In Fig.~\ref{fig3}  spectra of $D_{\rm ov}$ 
are shown
for various lattice spacings and a quark mass of $m=0.2$ with the index
of the Wilson kernel equal to $\nu =1$. After rescaling the eigenvalues,
the curves around the origin  collapse onto a single curve outside the Aoki phase
and to a different curve inside this phase  
(see Fig.~\ref{fig3}). 
The reason is that, in the Aoki phase, the ensemble
is a superposition of 
configurations with 0 or 1 zero modes with a very small admixture of
 more zero modes. Hence, the spectral density is a superposition of 
$\nu=0$ and $\nu=1$  Dirac spectra with each of them given by the
continuum chRMT result.

The agreement of the spectral density with  continuum chiral RMT (chRMT)
for the values of $a$ outside the Aoki phase can better be seen after unfolding the spectrum, see Fig.~\ref{fig2}. Thereby, the spectrum is transformed to have a constant  average level spacing  between consecutive levels.

\section{Conclusions}

Wilson chiral perturbation theory and the corresponding Wilson RMT are powerful
tools to study discretization effects of the Wilson Dirac operator. In the microscopic domain
as well as in the mean field limit this allows us to derive exact results for the spectra
of the Wilson Dirac operator and the Hermitian Wilson Dirac operator. 
Among others, the topological eigenvalues of the Dirac operator are 
broadened with a width that for small $a$ scales as $a /\sqrt V$. The width of the spectrum $D_{\rm W}$ is
directly related to the low-energy constants. For small lattice spacing we have derived
simple relations between properties of Dirac spectra and low energy constants. In this lecture,
we have discussed the mean field result for the Wilson Dirac spectrum with dynamical quarks and
have argued that the Dirac spectrum shows a first order jump for 
$W_8 +2 W_6 < 0$.
  This jump is induced
by the fermion determinant and only occurs in the presence of dynamical quarks.

Since lattice theories with the overlap Dirac operator preserve a 
form of chiral symmetry at  
nonzero lattice spacing we expect that the eigenvalues behave as in the continuum. Indeed, after unfolding, 
the correlations of the overlap Dirac operator are given by   continuum chRMT.
This is robust as long as the Wilson Dirac operator in the kernel is outside the Aoki phase. Inside the Aoki phase, we find a superposition of chRMT spectra
determined by the distribution of the zero modes.

\vspace*{1cm}
{\bf Acknowledgments.}
This work was supported by   U.S. DOE Grant No. DE-FAG-88FR40388 (JV and SV), the
Humboldt Foundation (MK) and 
the {\sl Sapere Aude} program of The Danish Council for 
Independent Research (KS).  
 We thank Gernot Akemann and Poul Damgaard  for fruitful discussions.

\end{document}